\documentclass[sn-mathphys-ay]{sn-jnl_arXiv}
\newcommand\arxiv[1]{#1} 

\usepackage[table]{xcolor}

\usepackage{placeins}
\usepackage{graphicx}%
\usepackage{multirow}%
\usepackage{amsmath,amssymb,amsfonts}%
\usepackage{amsthm}%
\usepackage{mathrsfs}%
\usepackage[title]{appendix}%
\usepackage{xcolor}%
\usepackage{textcomp}%
\usepackage{manyfoot}%
\usepackage{booktabs}%
\usepackage{algorithm}%
\usepackage{algorithmicx}%
\usepackage{algpseudocode}%
\usepackage{listings}%

\usepackage{microtype}
\usepackage{hyperref}
\usepackage{url}
\usepackage{booktabs}
\usepackage{comment}
\usepackage{tcolorbox}
\tcbuselibrary{skins}
\usepackage{ amssymb }
\usepackage{stackengine}
\usepackage{scalerel}
\usepackage{enumitem}
\usepackage{wasysym}
\usepackage{tikz}
\usetikzlibrary{svg.path}
\definecolor{lightblue}{RGB}{173,216,230}
\definecolor{mediumblue}{RGB}{0,0,205}
\definecolor{darkblue}{RGB}{0,0,139}

\definecolor{middarkred}{RGB}{204,0,0}
\definecolor{middarkorange}{RGB}{230,145,56}
\definecolor{middarkyellow}{RGB}{241,177,50}
\definecolor{middarkgreen}{RGB}{106,168,79}
\definecolor{middarkblue}{RGB}{61,133,198}

\definecolor{pastelred}{RGB}{255,179,186}
\definecolor{pastelorange}{RGB}{255,223,186}
\definecolor{pastelyellow}{RGB}{255,255,186}
\definecolor{pastelgreen}{RGB}{186,255,201}
\definecolor{pastelblue}{RGB}{186,225,255}

\hypersetup{
    colorlinks,
    linkcolor={red!50!black},
    citecolor={blue!50!black},
    urlcolor={blue!80!black}
}

\usepackage{tikz}

\newcommand{\easy}{%
  \item[{%
    \begin{tikzpicture}[x=1pt,y=1pt,scale=0.4,transform shape]
      \path[fill=middarkblue] 
        svg[yscale=-1]{M12 17q.825 0 1.413-.587T14 15t-.587-1.412T12 13t-1.412.588T10 15t.588 1.413T12 17m-6 5q-.825 0-1.412-.587T4 20V10q0-.825.588-1.412T6 8h7V6q0-2.075 1.463-3.537T18 1q1.875 0 3.263 1.213T22.925 5.2q.05.325-.225.563T22 6t-.7-.175t-.4-.575q-.275-.95-1.062-1.6T18 3q-1.25 0-2.125.875T15 6v2h3q.825 0 1.413.588T20 10v10q0 .825-.587 1.413T18 22z};
    \end{tikzpicture}%
  }]%
}

\newcommand{\medium}{%
  \item[{%
    \begin{tikzpicture}[x=1pt,y=1pt,scale=0.55,transform shape]
      \path[fill=middarkyellow] 
        svg[yscale=-1]{M5 3.5a3 3 0 0 1 6 0V4h.5a2.5 2.5 0 0 1 2.471 2.119a4 4 0 0 0-4.934 4.43L6.586 13a2 2 0 0 0-.543 1H4.5A2.5 2.5 0 0 1 2 11.5v-5A2.5 2.5 0 0 1 4.5 4H5zm3-2a2 2 0 0 0-2 2V4h4v-.5a2 2 0 0 0-2-2M8 10a1 1 0 1 0 0-2a1 1 0 0 0 0 2m5 3a3 3 0 1 0-2.871-2.129l-2.836 2.836a1 1 0 0 0-.293.707V15.5a.5.5 0 0 0 .5.5h2a.5.5 0 0 0 .5-.5V15h.5a.5.5 0 0 0 .5-.5V14h.5a.5.5 0 0 0 .5-.5v-.67c.313.11.65.17 1 .17m.75-4.5a.75.75 0 1 1 0 1.5a.75.75 0 0 1 0-1.5};
    \end{tikzpicture}%
  }]%
}

\newcommand{\hard}{%
  \item[{%
    \begin{tikzpicture}[x=1pt,y=1pt,scale=0.4,transform shape]
      \path[fill=middarkred] 
        svg[yscale=-1]{M18 8h-1V6c0-2.76-2.24-5-5-5S7 3.24 7 6v2H6c-1.1 0-2 .9-2 2v10c0 1.1.9 2 2 2h12c1.1 0 2-.9 2-2V10c0-1.1-.9-2-2-2m-6 9c-1.1 0-2-.9-2-2s.9-2 2-2s2 .9 2 2s-.9 2-2 2M9 8V6c0-1.66 1.34-3 3-3s3 1.34 3 3v2z};
    \end{tikzpicture}%
  }]%
}

\begin{document}

\title[Open Human Feedback]{The Future of Open Human Feedback}



\author[1]{\fnm{Shachar} \sur{Don-Yehiya}}

\author[2]{\fnm{Ben} \sur{Burtenshaw}}

\author[3]{\fnm{Ramon} \sur{Fernandez Astudillo}}

\author[4]{\fnm{Cailean} \sur{Osborne}}

\author[5]{\fnm{Mimansa} \sur{Jaiswal}}

\author[6]{\fnm{Tzu-Sheng} \sur{Kuo}}

\author[7]{\fnm{Wenting} \sur{Zhao}}

\author[8]{\fnm{Idan} \sur{Shenfeld}}

\author[8]{\fnm{Andi} \sur{Peng}}

\author[9]{\fnm{Mikhail} \sur{Yurochkin}}

\author[10]{\fnm{Atoosa} \sur{Kasirzadeh}}

\author[11]{\fnm{Yangsibo} \sur{Huang}}

\author[12]{\fnm{Tatsunori}\sur{Hashimoto}}

\author[2]{\fnm{Yacine} \sur{Jernite}}

\author[2]{\fnm{Daniel} \sur{Vila-Suero}}

\author[1]{\fnm{Omri} \sur{Abend}}



\author[13]{\fnm{Jennifer} \sur{Ding}}

\author[14]{\fnm{Sara} \sur{Hooker}}

\author*[4]{\fnm{Hannah Rose} \sur{Kirk}}\email{hannah.kirk@oii.ox.ac.uk}

\author*[8,9]{\fnm{Leshem} \sur{Choshen}}\email{leshem.choshen@ibm.com}

\affil[1]{\orgname{Hebrew University}}

\affil[2]{\orgname{Hugging Face}}
\affil[3]{\orgname{IBM}}

\affil[4]{\orgname{University of Oxford}}

\affil[5]{\orgname{Independent Researcher}}

\affil[6]{\orgname{CMU}}

\affil[7]{\orgname{Cornell University}}

\affil[8]{\orgname{MIT}}

\affil[9]{\orgname{MIT-IBM Watson AI Lab}}

\affil[10]{\orgname{University of Edinburgh}}

\affil[11]{\orgname{Princeton University}}

\affil[12]{\orgname{Stanford University}}

\affil[13]{\orgname{The Alan Turing Institute}}

\affil[14]{\orgname{Cohere For AI}}

%


\raggedbottom

\abstract{
Human feedback on conversations with language language models (LLMs) is central to how these systems learn about the world, improve their capabilities, and are steered toward desirable and safe behaviors. However, this feedback is mostly collected by frontier AI labs and kept behind closed doors. In this work, we bring together interdisciplinary experts to assess the opportunities and challenges to realizing an open ecosystem of human feedback for AI. We first look for successful practices in peer production,  open source, and citizen science communities. We then characterize the main challenges for open human feedback. For each, we survey current approaches and offer recommendations. 
We end by envisioning the components needed to underpin a sustainable and open human feedback ecosystem. In the center of this ecosystem are mutually beneficial feedback loops, between users and specialized models, incentivizing a diverse stakeholders community of model trainers and feedback providers to support a general open feedback pool. 




}
\keywords{Language Model, Open Models and Data, Feedback, RLHF}

\maketitle

\section{Introduction}
Natural language conversations (i.e., chats) have become a primary mode of human-AI interaction. Since the debut of large language models (LLMs) as general-purpose, multi-task reasoners \citep{alec2019language}, the volume and variety of use cases have grown, from providing information \citep{ivanova2024elements} and enabling tool use \citep{schick2024toolformer} to  assisting with writing text \citep{imran2023analyzing} and code \citep{barke2023grounded}. The steerability and usability of modern LLMs are underpinned by \emph{fine-tuning alignment} through human feedback (e.g., preferences), which guides LLM outputs toward desirable properties such as helpfulness, informativeness, and safety \citep[e.g., ][]{askellGeneral2021,ouyangTraining2022, dang2024rlhfspeaklanguagesunlocking,bai2022training, thoppilan2022lamda, nakanoWebGPT2021,wang2024helpsteer2,ahmadian2024basicsrevisitingreinforcestyle}. Despite being a cornerstone of modern AI research and development (R\&D), 
mechanisms for sustainably sourcing and sharing human feedback data are still underdeveloped for at least three reasons.

Firstly, many high-quality human preference datasets are proprietary and hence not available for others to study, reuse, or modify \citep{patel2023google}. This makes it difficult to infer best practices in the structure and type of preference data that leads to the largest gains in performance \citep{metaaiIntroducing2024}. Moreover, often the methodologies are not released, preventing even replication and further study (see Table~\S\ref{tab:open_datasets}).

Secondly, many barriers prevent sharing new datasets, in particular the cost and effort needed for extensive human annotation \citep{boubdir2023promptsmakedifferencedata}.  
For example, feedback collection often requires sophisticated user interface (UI) design \citep{singh2024aya} and lengthy ethics reviews. The costs further compound when specialist expertise is required
\citep{liWMDP2024}. The combination of demonstrable value and expensive collection results in dearth of open feedback datasets.

Third, unlike many ongoing feedback collection efforts of frontier AI labs \citep{LLaMa31,ouyang2022training}, open feedback datasets are often treated as static collections of uniform preferences rather than living artifacts that are dynamically maintained and updated.  
There are growing efforts to circumvent the costs of dedicated feedback collection. Some widely-used datasets instead collate implicit feedback from existing platform interactions such as on Reddit \citep{stiennonLearning2020} and Stack Overflow \citep{lambertHuggingFace2023};
or an increasingly popular loophole is to simulate feedback using powerful closed models \citep{cuiUltraFeedback2023,alpaca,aakanksha2024multilingualalignmentprismaligning}. However, distillation of one model's outputs to improve another carries legal, reproducibility, and transparency issues and adds a dependency on closed-model ecosystems. 
Comparatively, efforts that directly interface with real human-model interactions are rare. These efforts often base their success on hosting LLMs that users can freely use in exchange for chat logs \citep{zhao2024wildchat} or preference data \citep{zhengJudging2023,singh2024aya}; though other mechanisms of exchange exist, including paid participation in large-scale human studies \citep{kirkPRISM2024, aroyoDICES2023}, or mediator-enabled donation schemes where users retain ownership \citep{ShareLMS76:online}. These contributions are already transforming the R\&D of open models, but have known obstacles to sustainability such as depending on a small group of self-selected annotators \citep{zhao2024wildchat, zhengJudging2023} and being skewed towards bulk contributors \citep{kopf2024openassistant, singh2024aya}. Additionally, efforts that attempt to fairly remunerate contributors depend on funding \citep{kirkPRISM2024}.



Given these challenges, an unresolved question remains: \textbf{\textit{How can we develop and maintain sustainable systems of open human feedback for current and future eras of AI R\&D?}} 
In this work, we seek to encourage progress by presenting the key components required for a sustainable and open human feedback ecosystem. We solicit inputs from interdisciplinary experts across academic, industry, and open source communities. Our work consists of three main contributions:

\begin{enumerate}

    \item We draw from best practices of peer production and open source software (OSS) communities to identify shared requirements conducive to LLM development (\S\ref{sec:background}). 
    \item  We present seven themes that we view as central to the future of open human feedback (\S\ref{sec:challenges} \& \hyperlink{summary}{Summary}). For each theme, we summarize why it is an important challenge, potential existing solutions, and recommendations for building a more sustainable model of open human feedback.
    \item We conclude by examining what components underpin a successful ecosystem to support further progress (\S\ref{sec:roadmap}). 
    We acknowledge that operationalizing sustainability in practice will require trade-offs and difficult prioritization between these themes and substantial buy-in from various stakeholders. To address these challenges, we propose forming an ecosystem based on effective platforms, aligned private-communal incentives, and feedback loops for specialized communities (\S\ref{fig:figure-1-ecosystem}).
\end{enumerate}


\subsection{Defining Open Human Feedback}
\label{sec:definition}

We begin by defining our scope, clarifying what we mean by \textit{open}, \textit{human}, and \textit{feedback}.

By \textit{human}, we require some degree of human intervention and judgment to the extent that the feedback is not fully automatically generated, such as models simulating human feedback \citep{cui2023ultrafeedback, agnew2024illusion}. This does not exclude aided or seeded feedback (e.g., see platforms \S\ref{sec:roadmap} and effort \S\ref{subsec:effort}). 

By \textit{feedback}, we mean human responses to model outputs that are ascribed with some positive or negative valence, i.e., a value judgement.
Moreover, human feedback must comment on \textit{model} outputs. For example, traditional labeled datasets are not a form of feedback, as they are annotated by humans over human generated text and hence contain no reaction element. Another edge case is ``wizard-of-oz'' datasets \citep{kopf2024openassistant}, where humans pretend to be models, though these are rare.

Clear terminology surrounding ``open'' versus ``closed'' feedback is also required, especially given growing concerns about open-washing in ``open source AI'', despite substantial differences in transparency, accessibility, and governance structures \citep[e.g.][]{white2024model,liesenfeld2024rethinking}. We consider openness along five non-binary axes (see current datasets in Table~\S\ref{tab:open_datasets}): 
\begin{enumerate}
 \item \textbf{Open methodology:} Whether sufficient information is given to reproduce the dataset, at least partially (e.g., annotators details, guidelines and interfaces).
 \item \textbf{Open access:} Whether the final dataset is released for public access and use under a permissive license.
 \item \textbf{Open model participation:} Whether all feedback is collected from one predetermined LLM or if third-parties can upload and include their models.
 \item \textbf{Open human participation:} Whether feedback is provided by diverse annotators representing the views of LLM users, in contrast to crowdworkers or hired contractors who may enforce narrow views of ideal model behavior. 
 \item \textbf{Open timeline:} Whether feedback is dynamic, covering new models, new topics and capabilities, or is a one-time effort with a short or single time frame. 
\end{enumerate} 

Combining these elements, we consider \textbf{\textit{open human feedback}} \textit{as human-generated, valenced responses to AI model outputs characterized by open data accessibility and permissive terms of use, with ongoing and inclusive participation by humans and AI models.}


\newcommand{\Yes}{\cellcolor{green!25} Yes}
\newcommand{\No}{\cellcolor{red!25} No}

\begin{table}
\centering
\footnotesize
 \resizebox{\textwidth}{!}{%
\begin{tabular}{lcccccc}
\toprule
 & Source & Methodology & Access & Model & Human & Timeline \\
  &  &  & & Participation & Participation &  \\
\midrule
ShareLM &\citep{ShareLMS76:online} & \Yes  & \Yes & \Yes & \Yes & \Yes \\
Chatbot Arena & \citep{zheng2024lmsyschatm} & \Yes  & \Yes$^{\dagger}$ & \Yes & \Yes & \Yes \\
WildChat & \citep{zhao2024wildchat} & \Yes  & \Yes & \No  & \Yes & \Yes \\
PRISM & \citep{kirkPRISM2024}  & \Yes      & \Yes & \No & \Yes & \No \\
HelpSteer2 & \citep{wang2024helpsteer2}  & \Yes & \Yes & \No & \No & \No \\
DICES & \citep{aroyoDICES2023}  & \Yes & \Yes & \No & \No & \No \\
AnthropicHH & \citep{bai2022training}  & \Yes      & \Yes & \No & \No & \No \\
WebGPT & \citep{nakanoWebGPT2021} & \Yes  & \Yes & \No & \No & \No \\
LLaMa 3.1 &\citep{LLaMa31}  & \Yes   & \No & \No & \No & \No \\
LLaMa 2 & \citep{touvron2023llama}    & \Yes    & \No & \No & \No & \No \\
InstructGPT & \citep{ouyangTraining2022}  & \Yes   & \No & \No & \No & \No \\
GPT4 & \citep{achiam2023gpt}  & \No   & \No & \No & \No & \No \\
Claude 2 &\citep{claude2}  & \No   & \No & \No & \No & \No \\
Claude 3 &\citep{claude3}  & \No   & \No & \No & \No & \No \\
Gemini &\citep{team2023gemini}  & \No   & \No & \No & \No & \No \\ 
Gemini 1.5 &\citep{reid2024gemini}  & \No   & \No & \No & \No & \No \\ 
\end{tabular}
}
\caption{Human feedback datasets (or articles mentioning closed human feedback data) categorised for whether they are open along various dimensions we consider. $^{\dagger}$ the largest volume of prompts and feedback remains unpublished. \label{tab:open_datasets}}
\end{table}



\section{Key Lessons from Peer Production and Open Source}\label{sec:background}

Open feedback for LLMs can draw inspiration from the practices of peer production and OSS communities in promoting open, sustained, and diverse contributions. We provide an overview of these communities and highlight key takeaways from their success.

\subsection{Peer Production}\label{sec:governance}

Peer production is a (knowledge) production model in which numerous individuals collaborate, often remotely, without relying on traditional hierarchical organization or financial compensation \citep{benkler2007wealth}. Peer production communities have effectively brought together diverse contributors (see \S\ref{sec:diversity}) to pursue their shared missions. For example, Wikipedians
make 290k edits per day to create and maintain a free online encyclopedia \citep{halfaker2020ores} and OpenStreetMap enthusiastic mappers maintain open digital maps all over the world \citep{palen2015success}.

The success of peer production projects relies on both intrinsically and extrinsically motivated contributors. For example, volunteers contribute to Wikipedia because they believe in the shared mission and fundamental value of free online encyclopedias for humanity \citep{bryant2005becoming}, but they may also contribute for other reasons such as intellectual stimulation and enjoyment \citep{balestra2017investigating}. Wikipedia's Barnstars \citep{kriplean2008articulations} and Stack Overflow's reputation scores \citep{mamykina2011design,movshovitz2013analysis} are examples that have fostered external motivations
. Sometimes, extrinsic motivators can backfire; for example, psychological research notes that  when extrinsic motivations are introduced existing intrinsic motivations may dissipate \citep{deci1971effects, ryan2020intrinsic}.

Another factor contributing to the success and sustainability of peer production is the availability of platforms that provide the infrastructure, processes, and tools that support community-driven governance \citep{heltweg2023}. For example, Meta Stack Overflow allows the Stack Overflow communities to discuss policies and suggest new features \citep{fang2023people}. Wikipedia's infrastructure enables the creation of policies, guidelines, and essays for community governance \citep{butler2008don}. Flexible governance is especially critical, as each community has unique governance needs \citep{zuckerman2023community}; even within Wikipedia, different language editions have unique rules \citep{hwang2022rules}.
Community-driven governance is critical for sustained contribution and long-term community health. Open human feedback should follow suit in supporting bottom-up, community-driven governance for sustainably obtaining and moderating high-quality data \citep{kuo2024wikibench}.

Forms of peer production in AI span a wide range of organization structures and focus areas. Some organizations are structured primarily around the production and exchange of knowledge on AI systems or data in specific linguistic and geographical contexts \citep{Masakhane, singh2024aya}, category of model architectures \citep{Peng2023RWKVRR}, a specific model \citep{bloom}, or broader focus on values, such as accessibility and replicability \citep{pmlr-v202-biderman23a}. These efforts have unique perspectives on collaboration, governance, code, data, and more~\citep{ding2023openness}. In most cases, however, explicit statements of values~\citep{ethical-charters}, governance processes for the project and artefacts~\citep{hughes2023bigcode}, and codes of conduct \citep{singh2024aya,Peng2023RWKVRR} play an important role in ensuring that the work benefits the targeted communities and, thus, motivates  contributions.


\subsection{Open source software}
\label{sec:background:oss}

OSS concerns the development and distribution of software under open source licenses that allow anyone to inspect, use, modify, or redistribute the source code \citep{osi_open_2007}. This modus operandi has enabled countless research and innovation achievements, including in AI \citep{langenkamp2022open,osborne2024ai,ding2023openness}.

OSS projects are developed by diverse contributors, driven by varied social, economic, and technological incentives \citep{feller2002understanding, bonaccorsi2006comparing}. For individuals, intrinsic incentives, such as personal values, needs, or enjoyment, as well as extrinsic incentives, such as reputation, learning, and career benefits, are significant \citep{von2012carrots,lakhani2005hackers}. Individuals' incentives vary based on their volunteer vs. paid status \citep{lakhani2005hackers}, project governance \citep{shah2006motivation}, and their cultural norms \citep{subramanyam2008free,takhteyev2012coding}. 

Companies are chiefly driven by economic and technological incentives \citep{bonaccorsi2006comparing,li2024systematic}, such as reducing R\&D costs \citep{lindman2009beyond, birkinbine2020incorporating}, shaping standards \citep{fink2003business,lerner2002some}, practicing reciprocity \citep{osborne2024publicprivate,pitt2006penguin}, and recruiting talent \citep{aagerfalk2008outsourcing, west2006challenges}, among others. 

A key lesson from successful OSS projects is the importance of open governance and vendor-neutral hosting (e.g., by an independent consortium or foundation), which facilitate contributors with diverse incentives to build consensus \citep{o2008boundary} and collaborate on non-differentiating base-technologies \citep{germonprez2013open}. Similarly to peer production communities, OSS communities have created tools that enable open governance and collaboration, such as standard licenses, codes of conduct, and tools and metrics that can support monitoring and improving project and ecosystem health over time \citep{goggins2021open}. These practices can similarly ensure broad participation and sustained innovation in the open human feedback ecosystem.


\section{Challenges and Opportunities of Realizing Open Human Feedback}\label{sec:challenges}

We  present seven themes that are central to the future of open human feedback. For each theme, we summarize why it is an important challenge, potential existing solutions, and recommendations for building a sustainable ecosystem for open human feedback.


\subsection{Incentives}\label{sec:incentives}

Aligned incentives are needed to motivate individuals and companies to participate in sharing or annotating open human feedback data. As per \S\ref{sec:background}, we must account and design for the diverse incentives of contributors, from individuals to companies. 



With regards to individuals, the open human feedback ecosystem ought to accommodate both intrinsic and extrinsic incentives of diverse individuals, from volunteers to sponsored contributors. Intrinsic incentives may include learning and skill development, hobbyism, and community kinship, among others. For example, contributors from underrepresented socio-demographic and/or socio-linguistic groups might participate in open human feedback projects to advance AI resources and alignment for the interests and needs of their community \citep{singh2024aya,pipatanakul2023typhoon}. It is advised that projects establish  governance structures (including a mission statement) that resonate with their community members' values. However, the ``burden'' of participation needs careful co-design to mitigate exploitation, benefiting solely those who build and commercialize AI systems \citep{kuo2024wikibench, birhane2022power, sloane2022participation}. For extrinsic incentives, projects could encourage participation via leaderboards, gamification points, or monetary compensation \citep{krishnamurthy2006bounty}. 

With regards to companies, we must distinguish between frontier AI companies that would share human feedback data and companies that would contribute to open human feedback data projects. First, frontier AI companies can be incentivized to share human feedback collected via their proprietary chatbots by aligning data sharing with their strategic interests, such as reducing R\&D costs (e.g., via crowdsourcing annotations) and building recognition in the open source AI community (e.g., via leaderboards). Examples of successfully spun-out OSS and open data projects may help temper concerns and illustrate the benefits of sharing data for open collaboration. In order to facilitate contributions by and collaboration among diverse companies to human feedback data projects, we call for such projects to be hosted by a non-profit, vendor-neutral organization (such as the \href{https://humanfeedback.io/}{Human Feedback Foundation}), rather than by a single company, as a tried-and-tested enabler of collaboration between ``unexpected allies'' \citep{o2008boundary,germonprez2013open}. Foundations with relevant expertise and missions include the \href{https://humanfeedback.io/}{Human Feedback Foundation}, \href{https://genaicommons.org/}{GenAI Commons}, and the \href{https://lfaidata.foundation/}{LF AI \& Data Foundation}. 



\subsection{Effort and Involvement}\label{subsec:effort}


The success of the open human feedback ecosystem will rely on substantial human effort involving a range of tasks, such as gathering samples, annotating data, and writing alternative answers. Given the range, platforms intended for feedback collection should provide simple interfaces that reduce cognitive load and aid task completion \citep{chen2011eye, ash2018digital}, which is especially important to sustain volunteer and decentralized workflows. At the same time, feedback collection mechanisms should be be implemented at a centralized level in a dedicated platform \citep{boubdir2023promptsmakedifferencedata}.
Most current feedback platforms use pairwise comparisons for ranking LLMs~\citep{zheng2024lmsyschatm}, but these comparisons typically lack fine-grained assessment (\S~\ref{sec:diversity}, \ref{sec:domain_specificity}). Moreover, feedback on these existing platforms tends to be of limited diversity and coverage.
The paired back functionality of these hosted platforms 
attracts short conversations and relatively low topical, usecase and user diversity \citep{lin2024wildbench}.
The simplest way to reduce unnecessary effort is to make intuitive user interfaces for feedback platforms.  Where possible, prompted feedback, such as users' rating or ranking  of LLM responses, should be supplemented with feedback from the naturally-occurring cues in the chat ~\citep{hancock-etal-2019-learning}, such as when the user verbally thanks the model or edits their original prompt \citep{donyehiya2024naturalfeedback}. 


\subsection{Expert Contributions}\label{sec:domain_specificity}


LLMs often perform poorly at tasks that require expert knowledge \citep{gougherty2024testing}, and specialized feedback from domains such as healthcare, legal, or finance is more challenging to acquire than general feedback. High-quality specialized feedback is unlikely to naturally occur through experts' use of LLMs, and hiring domain experts is costly. Nonetheless, expert feedback is necessary for evaluating model capabilities and guiding improvements, especially in complex or technical domains where layperson knowledge is not sufficient.
Current efforts to collect expert feedback are limited in size and diversity, and/or gate-kept within private organizations \citep{Pokrywka2024GPT4PM, MerlynEducationModels}. For example, GPQA~\citep{rein2023gpqa} collects static expert contributions, while Bloomberg GPT relies on internal company data \citep{wu2023bloomberggpt}. 
Companies like OpenAI or DeepMind employ specialized AI trainers and keep the proprietary feedback `in-house'~\citep{OpenAI_2024}.
Replicating these efforts in open source communities faces challenges including high costs (GPQA costs $\sim$\$100/hr), scalability issues \citep{liu2023fingpt}, and quality control. Moreover, these efforts risk domain bias; for example, AlpacaEval overrepresents coding while neglecting theoretical mathematics \citep{lin2024wildbench}.
Successful initiatives for driving expert participation like Wikipedia editing and OSS development (see \S\ref{sec:background}) allow contributors to choose their tasks based on their interests and skill~\citep{klie2023lessons}. To attract experts, platforms should provide tools for identifying topics requiring specialized input, including fact-checking and  deliberation, and for routing these tasks to well-suited contributors.
Different incentive mechanisms may appeal to experts, including contributing to  the development of specialized models or public attribution and recognition for their contributions.




\subsection{Linguistic and Cultural Diversity}\label{sec:diversity}

LLMs struggle to serve diverse human demographics due to limited coverage in their general and feedback training data, which predominately relies on English speakers from narrow communities \citep{zhao2024wildchat, zhengJudging2023, pavlick2014language}, and few annotators contribute the majority of data \citep{kopf2024openassistant}, even in multilingual efforts \citet{singh2024aya}. 
Accordingly, culturally-specific queries and minority perspectives tend to be overlooked \citep{zhao-etal-2024-uncommonsense}. 
To build technologies inclusive of diverse socio-demographic and socio-linguistic groups \citep{seth2024dosa,aakanksha2024multilingualalignmentprismaligning}, it is crucial to go beyond ``convenience samples'' \citep{emerson2015convenience} that rely solely on LLM developers, enthusiasts \citep{zheng2024lmsyschatm}, or company-employed annotators \citep{LLaMa31}, and to gather feedback from underrepresented communities \citep{watts2024pariksha, quaye2024adversarial}. However, reaching different communities can be challenging due to factors such as lack of technology access \citep{tsatsou2011digital} or different device uses, such as the common use of mobile in the Global South \citep{Avle2018Research}.

There have been few recent attempts at covering demographic and geographic depth \citep{kirkPRISM2024,singh2024aya}, or multilingualism and dialect diversity \citep{singh2024aya,lu2024llamaxscalinglinguistichorizons,aakanksha2024multilingualalignmentprismaligning,watts2024pariksha}. 
Community-led engagement can return more power and control over data to the communities themselves \citep{kuo2024wikibench, petersParticipation2018}, but must be handled with care to avoid exploitative practices \citep{birhane2022power, sloane2022participation, klie2023lessons}. 
Opening human feedback data to peer production provides the single most promising step for diverse participation. However, diversity is not guaranteed (\S\ref{sec:background}) and to facilitate it, platforms should be accessible, available in different languages, and be complemented by outreach to underrepresented groups to uncover their unfulfilled needs. Moreover, clear guidelines and regular audits \citep{santurkar2023whose} might reduce opinion biases and uncover representation gaps. 




\subsection{Adaptable and Dynamic Feedback}\label{sec:adaptable-and-up-to-date}


Human feedback differs from traditional dataset annotations: while resolving grammatical sentences or solving linear equations have a single, static ground truth; human preferences, especially those surrounding cultural issues, moral debates, and social conventions vary across individuals and groups (\S~\ref{sec:diversity}) as well as over time \citep{pozzobon2023goodtrieveradaptivetoxicitymitigation}. As human preferences and views differ and evolve, model outputs will necessitate ongoing feedback collection and models trained on existing feedback may require feedback on new types of errors.
This bears similarity to existing platforms for ongoing annotation. For example, Dynabench allows iterative model benchmarking \citep{kiela-etal-2021-dynabench}, LiveBench mines new examples \citep{white2024livebenchchallengingcontaminationfreellm}, ChatBot Arena publicises new model releases \citep{zheng2024lmsyschatm}, and the ShareLM browser plug-in mediates ongoing user conversations on many platforms \citep{ShareLMS76:online}.

Acquiring dynamic data is challenging and costly, rendering traditional payment for annotation, as used in static datasets, unsuitable. Instead, to achieve continuous participation from the community, longitudinal incentives need to be designed and built into the platform (\S\ref{sec:incentives}) and new methods to be invented to flag and update outdated feedback. In time, the dated feedback itself might become useful, for example in comparison with the updated one.

\subsection{Privacy and Data Protection}\label{sec:privacy}

Sharing feedback can have wide reaching benefits, including for contributors, but equally could harm the contributors if their privacy is not adequately protected. Feedback platforms might also motivate more contributions by upholding data protection legislation and adopting best practices that enable contributors to have a say over how their data is shared. Clear mechanisms of opt-out can support these aims, for example, the ``\textit{\href{https://huggingface.co/spaces/bigcode/in-the-stack}{Am I in The Stack?}''} initiative associated with the BigCode Dataset, which allows code developers to see whether their work is included in the open source software training dataset, and provides a mechanism for them to request to opt out their data from future versions of the dataset \citep{bigcodeAm2024}. Feedback platforms must comply with existing privacy regulations that standardize and enforce this social responsibility, such as GDPR \citep{EuropeanParliament2016a}, CCPA \citep{illman2019california}, and regulations for specific applications such as the Health \citep[HIPPA,][]{hipaa}. 

In addition to proper security protocols \citep{kumar2022towards}, anonymization and de-identification are central considerations prior to collective data sharing. This includes removing personally identifiable information (PII), such as names, emails, and IP addresses \citep{zhao2024wildchat, 10.1145/3351095.3372873}, which can be especially challenging when interleaved in a chat \citep{10.1145/3442188.3445922} and because PII definitions vary by application \citep{hipaa}. Furthermore, in feedback discussing specialized topics or rare dialects author identities might also be unavoidably compromised by the topic itself. 

Beyond anonymization, other privacy-enhancing techniques are also recommended by standard practices \citep[e.g.,][]{nistframework}. Moreover, the text itself inherently shares information that for example could be crossed with other sources to deanonymize \citep{narayanan2008robust}. Thus, unless privacy leakage is theoretically restricted -- e.g. by noisy aggregation of the data \citep{dwork2006calibrating, dwork2014algorithmic, cummings2023challenges} -- then research should at least estimate the privacy leakage.



Giving contributors control and autonomy over their personal information is vital for building trust and ensuring legal compliance. For example, allowing users to retract or edit their contributions, or providing a time window for complete removal of their data before feedback is made public \citep{ShareLMS76:online}. Privacy must be balanced with providing a sense of ownership (\S\ref{sec:legal}) as the two might be at odds. As heuristics like hashing names to match users to their content are susceptible to misuse, logins are often preferable \citep{hughes2023bigcode, ShareLMS76:online}. Regardless, any  opt-out guarantees presents logistical challenges as to how these requests would propagate to any derivative models trained on the data, which would then need to remove specific details \citep{DBLP:journals/corr/abs-2202-00885, tran2024measuring}, known as `machine unlearning' \citep{bourtoule2021machine, lynch2024eight, shi2024muse}.


\subsection{Legal and Ownership}\label{sec:legal}


To ensure all contributions are appropriately licensed and available for public use, while mitigating the risk of legal challenges, robust data and legal governance frameworks must be established and concisely communicated to contributors. 
Particularly, we propose following best licensing practices from successful open-source data collection efforts where contributors retain ownership of the data they share. Notably, both humans and models are involved in data creation, but commonly the creative act is legally attributed to the human \citep{guadamuz2017artificial}, we hence consider the data solely owned by the human contributor. This stance has precedents~\citep{openai_terms}, though the legal debate remains unsettled \citep{kop2020ai}. We suggest contributors are asked to give informed revocable consent to share their data under a permissive content license, such as Creative Commons or Open Data License~\citep{kim2007creative,zheng2024lmsyschatm}, by signing an agreement (e.g., checking a box). 
 This sets a transparent standard and reduces the fear of legal retribution for both the data collection platform and consumers of the data \citep{bonatti2017transparent}. 
Moreover, we suggest that those collecting feedback (e.g., platform providers) submit data card documentation \citep{
gebru2021datasheets, shimorina2021human, pushkarna2022data}, explaining context, methodology, intended applications, and limitations.



\section{Visions of Successful and Sustainable Open Feedback Ecosystems}\label{sec:roadmap}
Building on these challenges (see \hyperlink{summary}{Summary}), we conclude by considering the tangible components that are necessary for building an ecosystem of open human feedback under the mission: Humanity guiding open AI for humanity.

Any solution requires \textbf{building an open feedback platform}---sites or software that allow the community to dynamically contribute in a standardized and efficient way. Platform designs often shape standards for the community, and careful consideration is required to ensure they can facilitate efficient distribution of work (\S\ref{subsec:effort}), are accessible and easy to use by diverse contributors (\S\ref{sec:diversity}), enable governance and moderation of feedback (\S\ref{sec:governance}), and encourage participation (\S\ref{sec:incentives}). Beyond our current suggestions, such platforms should be open source and will evolve in line with changing feedback requirements, such as feedback types, biases, and privacy issues. While current platforms, especially Hugging Face's \href{https://huggingface.co/spaces}{Spaces}, have lowered the entry barrier for machine learning model experimentation, they lack detailed systematic feedback mechanisms. Industry-centered platforms, like ScaleAI, offer feedback platforms that support internal workforces to continuously refine models on datasets \citep{Iren_Bilgen_2014} but often suffer from biases due to limited diversity and specificity of crowdworker knowledge \citep[\S\ref{sec:diversity}][]{10.1145/3462204.3481729, 10.1145/3290605.3300773}. \href{https://argilla.io}{Argilla} 
represents the closest open source solution for feedback collection, yet it still lacks support for participants to discuss feedback, share incentives, or collectively govern projects.  

Another important component \citep{soumithICML} is a \textbf{shared resource} of the chats and feedback \citep{ShareLMS76:online}. Chats and feedback should be pooled together to simplify use. Ideally, this pool would  combine sustainable data from the platforms but also one-time efforts from the community \citep{kopf2024openassistant}. The mechanism of contributing to the pool might vary \citep{kuo2024wikibench}. At one extreme, chat UI platforms connect directly and share \citep{chiang2024chatbot, zhao2024wildchat}. Alternatively, sharing is done by dedicated mediators, like browser plug-ins\citep{ShareLMS76:online}. Mediators allow users to interact freely with models on any platform (even closed) and ensure the feedback is propagated to the pool. 


\begin{figure}[t]
    \centering
    \includegraphics[width= \textwidth]{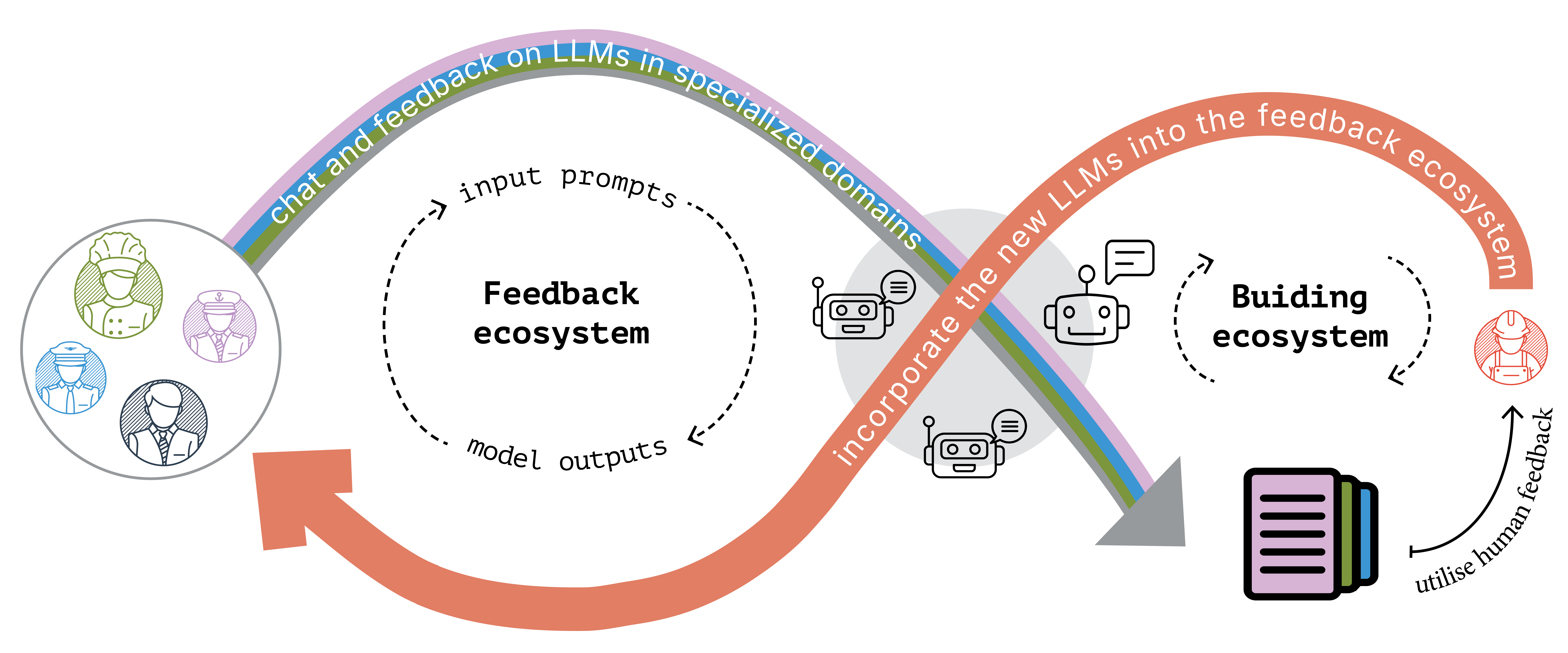}
    \caption{
    \textbf{A sustainable open feedback ecosystem at a glance}. 
    Users from diverse domains and expertise
    \includegraphics[height=1em]{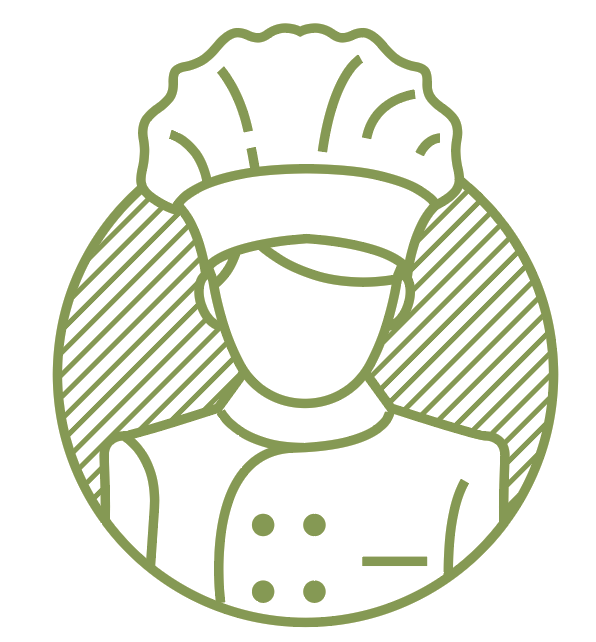}
    \includegraphics[height=1em]{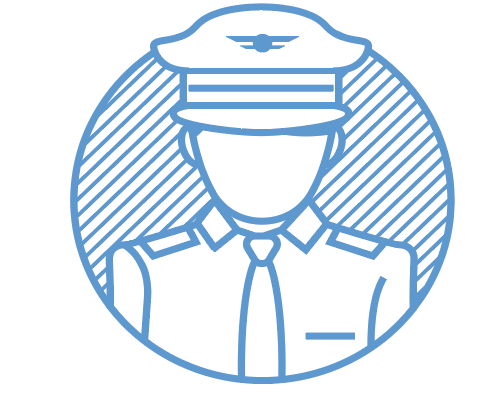}
    \includegraphics[height=1em]{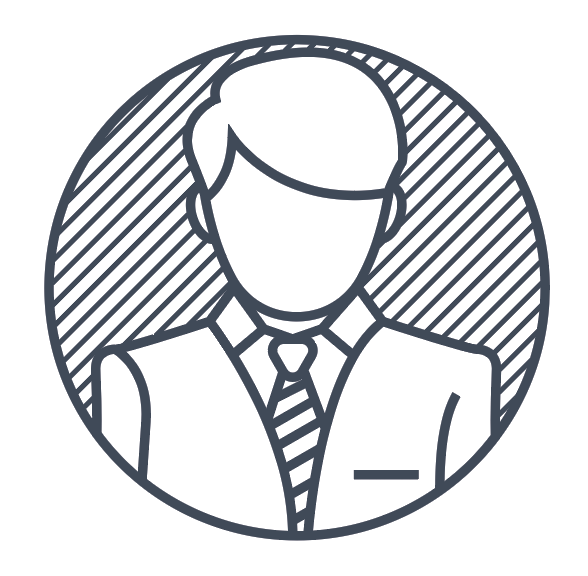}
    \includegraphics[height=1em]{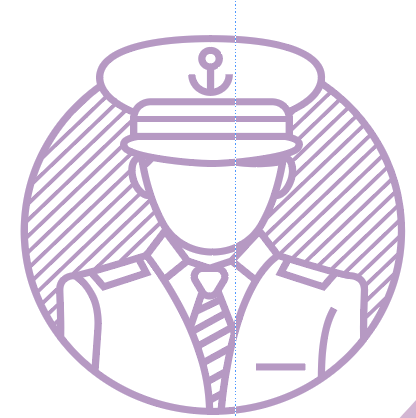}
    chat and give feedback to open Large Language Models within the feedback ecosystem
    \includegraphics[height=1em]{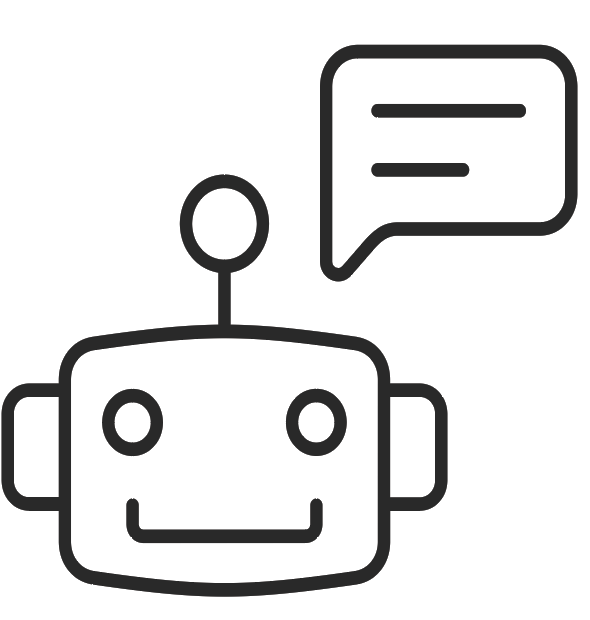}.
    The conversations and feedback are shared through an open pool of data
    \includegraphics[height=1em]{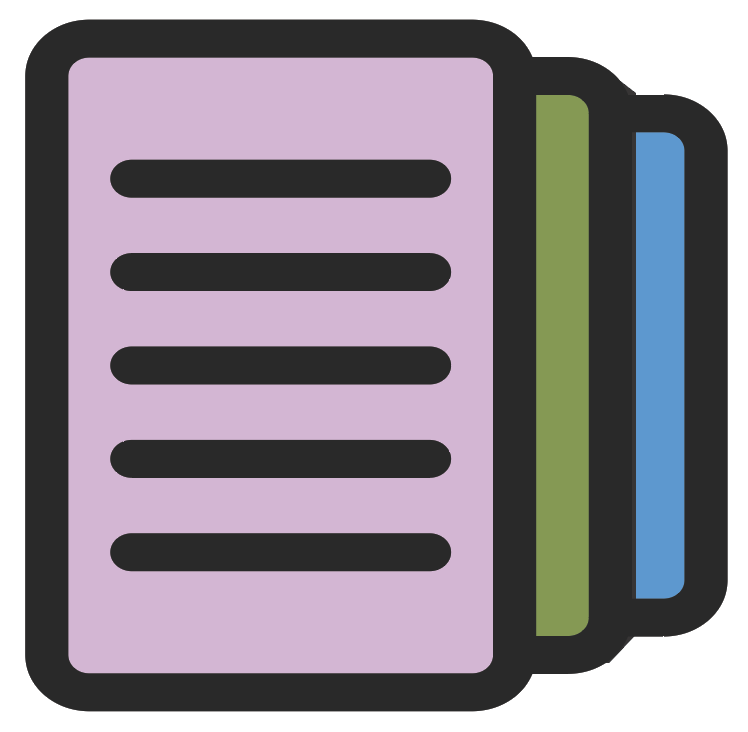},
    which can be utilized by AI builders in the building ecosystem 
    \includegraphics[height=1em]{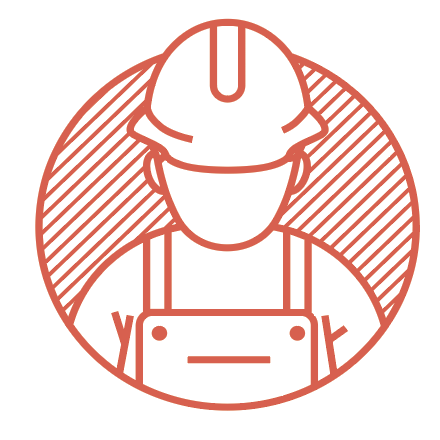}
    to develop and share the next generation of Large Language Models within the chat and feedback ecosystem.
    }
    \label{fig:figure-1-ecosystem}
\end{figure}

The last component is \textbf{participation}. To thrive, the ecosystem needs not only to improve the types and ways of collecting feedback, but to have a community. This requires aligning incentives of both individuals and organizational contributors of public resources, which underpins successful citizen science and peer production models \citep{nov2014scientists, chen2010social,bonaccorsi2006comparing}. 

To achieve all the requirements above (see \hyperlink{summary}{Summary}), we call for the open human feedback ecosystem to interoperate with the LLM training ecosystem (see Figure~\ref{fig:figure-1-ecosystem}). Specifically, \textbf{self-sustaining feedback loops}---systems of mutual benefit for all the stakeholders involved. Consider, for example, an open platform developed by the different contributors in the ecosystem. This platform spins up specialized models equipped with chat UI, model hosting, feedback collection, and automatic training on the acquired feedback, which specialize the model further. 
To further support the research community and its improvement, such a platform should provide functionalities for experimentation, adaptation, and adoption by new contributors, who may adapt feedback collection in custom ways that address their target questions. The platfom should also provide flexibility, allowing contributors to iteratively refine the system itself, observe the effect on feedback, and repeat, drawing on the success of the MAMA cycle (model, annotate, model, annotate) \citep{pustejovsky2012natural}.

Critically, the ecosystem will benefit contributors with specialized needs. For instance, consider a law student requiring detailed and accurate knowledge from an LLM for exam preparation. In a one-to-one relationship with a model, it would require substantial effort from the student alone to improve it for their needs. However, specialization can come from a communal feedback pool, and with time clustering or other unsupervised learning methods might offer personalization by identifying similar users in the overall community.
This process is analogous to how recommendation systems 
improve as more diverse users engage with the platform \citep{thorat2015survey}.

Consequently, a marketplace of models—specialized by topic, culture, or language—could be sustained if specialist contributors benefit from participating in these communities in exchange for their feedback contributions. This approach has the potential to create a continuously improving ecosystem, with specialized language models tailored to diverse user needs. Meanwhile, a standardized and central pool of feedback can serve the entire ecosystem. For example, the overall pooled data might be used to align more universal and generic models, to identify common patterns across domains, or to understand where irreconcilable differences between communities lie.

\section{Conclusion}
Human feedback has emerged as a central component for improving the capabilities and safety of AI models. However, human feedback data is concentrated within a few frontier AI companies, and there is currently no ecosystem for sharing and collaborating on such data. We have outlined our vision and key components required for a dynamic and sustainable open human feedback ecosystem: well-designed feedback platforms, aligned incentives for contributors, and feedback loops that can support specialized and general-purpose models. However, significant challenges remain: building and hosting feedback loops for specialized models, incentivizing diverse and quality contributions, protecting privacy, and managing governance and data ownership. Despite these hurdles, we call for recognition of open human feedback data as a critical, yet underdeveloped, ingredient for open and transparent AI R\&D, and propose a pathway forward.

\arxiv{\subsubsection*{Acknowledgments}
We thank Colin Raffel and Delip Rao for their helpful advice, comments, reviews and views during the document curation.}

\definecolor{main}{HTML}{13274F}
\begin{figure*}[thbp] \label{float:challenges}
\begin{tcolorbox}[width=\textwidth, colback=main!7, colframe=main, center title, fonttitle=\large\bfseries, title=\color{white} 
 \textbf{\hypertarget{summary}{Actions for Building the Open Human Feedback Ecosystem} } ]

\textbf{\hspace{0.5em} Incentivise participation  (\S\ref{sec:incentives})}
\vspace{0.1cm}
\begin{itemize}[leftmargin=*,itemindent=1em]
    \easy \parbox[t]{\dimexpr\linewidth-2em\relax}{State a clear mission that resonates with community values \\Proposed: Humanity guiding open AI for humanity}
    \easy Incentivize companies to share open human feedback data
    \easy Host projects in a vendor-neutral organization to facilitate participation
    \medium Design for intrinsic and extrinsic incentives of diverse contributors
\end{itemize}
\vspace{0.2cm}
\textbf{\hspace{0.5em} Reduce contribution barriers (\S\ref{subsec:effort})}
\vspace{0.1cm}
\begin{itemize}[leftmargin=*,itemindent=1em]
    \easy Design interfaces that reduce contribution barriers (e.g., time, effort, etc.)
    \medium Develop  efficient approaches to use, sample, and augment existing data
    \hard Extract feedback from natural text in chats
\end{itemize}
\vspace{0.2cm}
\textbf{\hspace{0.5em} Facilitate expert annotation  (\S\ref{sec:domain_specificity})}
\vspace{0.1cm}
\begin{itemize}[leftmargin=*,itemindent=1em]
    \medium Allow  the community to define feedback requirements independently 
    \medium Allow experts to choose what and how to contribute independently
\end{itemize}
\vspace{0.2cm}
\textbf{\hspace{0.5em} Diversify participation (\S\ref{sec:diversity})}
\vspace{0.1cm}
\begin{itemize}[leftmargin=*,itemindent=1em]
    \easy Facilitate peer-produced contributions to human feedback data projects
    \easy Encourage participation from diverse socio-demographic \& -linguistic groups
    \easy Mitigate exploitation through responsible use and contribution policies
    \easy Provide accessible and multilingual platform interface and tools 
    \medium Audit collected data for potential biases and discriminatory content
    \hard Design for feedback contestation and consensus-building among contributors
\end{itemize}
\vspace{0.2cm}
\textbf{\hspace{0.5em} Collect updated feedback (\S\ref{sec:adaptable-and-up-to-date})}
\vspace{0.1cm}
\begin{itemize}[leftmargin=*,itemindent=1em]
    \easy Pool data into one location
    \medium Make dedicated efforts for specialized data annotations
    \hard Create platforms, mediators, and tools for chat \& feedback data sharing
    \hard Encourage and enable the sharing of updated feedback data
    
\end{itemize}
\vspace{0.2cm}
\textbf{\hspace{0.5em} Privacy  (\S\ref{sec:privacy})}
\vspace{0.1cm}
\begin{itemize}[leftmargin=*,itemindent=1em]
    \easy Follow  security best practices and comply with relevant regulations
    \easy Map users to their contributions privately (e.g., via logins)
    \medium Develop methods to give users control over their personal information
    \hard Anonymize data and remove private information
\end{itemize}
\vspace{0.2cm}
\textbf{\hspace{0.5em} Legal (\S\ref{sec:legal})}
\vspace{0.1cm}
\begin{itemize}[leftmargin=*,itemindent=1em]
    \easy Share datasets with appropriate documentation (e.g., data card)
    \easy Ensure informed revocable consent for data sharing (e.g., via box-checking)
    \easy Users share data with permissive licenses (e.g.,  Creative Commons license)
\end{itemize}
\vspace{0.2cm}
\begin{tcolorbox}[
    enhanced,
    width=\textwidth,
    colframe=lightgray,
    boxrule=0.5pt,
    arc=2mm,
    title={},
    overlay={
        \node[fill=main!7, inner sep=2pt, anchor=west] 
        at ([xshift=10pt, yshift=0pt]frame.north west) 
        {\textbf{Legend}};
    }
]
\begin{itemize}[leftmargin=*,itemindent=1em]
    \easy Feasible with existing knowledge and tools
    \medium Knowledge and tools partially exist, but some R\&D still required
    \hard R\&D required
\end{itemize}
\end{tcolorbox}
\end{tcolorbox}
\end{figure*}

\FloatBarrier
\newpage
\bibliography{refs,anthology1,anthology2,HRKrefs}
\end{document}